\documentclass[aps,preprint,showpacs,groupedaddress,floatfix]{revtex4}

\usepackage{graphicx}
\usepackage{dcolumn}
\usepackage{bm}

\begin{document}

\bibliographystyle{prsty}
\author{N. Paar}
\affiliation{Physics Department, Faculty of Science, University of Zagreb, 
Croatia}
\title{Calculation of stellar electron-capture cross sections on nuclei based on microscopic Skyrme functionals}
\author{G. Col\`o}
\affiliation{Dipartimento di Fisica dell'Universit\`a degli Studi and INFN, Sezione di Milano,
via Celoria 16, 20133 Milano, Italy}
\author{E. Khan and D. Vretenar}
\affiliation{Institut de Physique Nucl\'eaire, IN$_{2}$P$_{3}$-CNRS/Universit\'e
Paris-Sud, 91406 Orsay, France}

\date{\today}
\begin{abstract}
A fully self-consistent microscopic framework for evaluation of nuclear
weak-interaction rates at finite temperature is introduced, based on Skyrme functionals. 
The single-nucleon basis and the corresponding thermal occupation factors of the initial nuclear 
state are determined in the finite-temperature Skyrme Hartree-Fock model, and 
charge-exchange transitions to excited states are computed using the 
finite-temperature RPA.  Effective interactions are implemented self-consistently: 
both the finite-temperature single-nucleon Hartree-Fock equations and the 
matrix equations of RPA are based on the same Skyrme energy density functional. 
Using a representative set of Skyrme functionals, the model is applied in the calculation 
of stellar electron-capture cross sections for selected nuclei in the iron mass group 
and for neutron-rich Ge isotopes.
\end{abstract}
\maketitle
\date{today}
\section{Introduction}

Nuclear weak-interaction processes play a crucial role in the late stages of evolution of
a massive star and in the pre-supernova stellar collapse \cite{Bet.90,LM.RMP03,Jan.07}. 
The core of a massive star at the end of hydrostatic burning is stabilized by electron 
degeneracy pressure, as long as its mass does not exceed the Chandrasekhar limit of 
about 1.44 solar masses. If this mass limit is exceeded, electron pressure can no longer stabilize 
the core and it collapses. The dynamics of this process depends on the 
core entropy and lepton-to-baryon ratio $Y_e$ \cite{BBAL.79}, which are essentially determined 
by weak-interaction processes: the nuclear $\beta$-decay 
\begin{equation}
({Z},{N}) \longrightarrow   ({Z+1},{N-1}) + e^{-} + \bar{ \nu}_{e}  \; ,
\end{equation}
and electron capture:
\begin{equation}
e^{-} + ({Z},{N}) \longrightarrow  \nu_{e} + ({Z-1},{N+1}) \; .
\end{equation}
The latter reduces the number of electrons available for pressure support, 
whereas beta decay acts in the opposite direction. At low matter densities 
$\rho \leq 10^{11}$ g cm$^{-3}$ neutrinos escape from the star, carrying away 
energy, i.e. cooling the stellar core, and keeping its entropy low. For initial $Y_e$
values of $\approx$ 0.5, $\beta$-decays are impeded by the presence of electrons 
which reduce the available phase space for decay, but become competitive when 
the composition of nuclei in the core gets more neutron-rich. In the early stage of 
the collapse, for densities lower 
than a few $10^{10}$ g cm$^{-3}$, the electron chemical potential is of the same order 
of magnitude as the nuclear Q value, and the electron-capture cross sections are 
sensitive to the details of the Gamow-Teller GT$^{+}$ strength distributions in 
daughter nuclei. At these densities and temperatures between 300 keV and 
800 keV, electrons are captured on nuclei with mass number $A \leq 60$. 
For higher densities and temperatures $T \approx 1$ MeV, electron capture occurs 
on heavier nuclei $A > 65$. Under these conditions, however, the electron chemical 
potential is significantly larger than the nuclear Q-value, and the capture rates are 
dominantly determined by the centroid and the total GT$^{+}$ strength \cite{LM.RMP03,Jan.07}. 

The first standard tabulation of nuclear weak-interaction rates for astrophysical applications 
was that of Fuller, Fowler and Newman \cite{FFN}. It was based on the independent particle model, 
but used experimental information whenever available. The tables included
rates for electron capture, positron capture, $\beta$-decay, and positron emission 
for relevant nuclei in the mass range $21 \leq A \leq 60$. Based on data which 
in the meantime became available on GT$^{+}$ strength distributions, and 
using large-scale shell-model diagonalization in the complete $pf$-shell, these
rates have been improved, and rates for electron and 
positron captures, and for $\beta^+$ and $\beta^-$ decays have been computed for 
stellar conditions and for more than 100 nuclei in the mass range $A = 45-65$ \cite{LM.00,Mar.00}.  
Using the improved weak-interaction rates, models for pre-supernova evolution of massive 
stars were examined in Ref.~\cite{Heg.01}, and it was concluded that the resulting changes 
in the initial value of $Y_e$ and iron core mass could have important consequences for 
nucleosynthesis and the supernova explosion mechanism. Detailed calculations of stellar 
weak-interaction rates in the iron mass region have also been carried out with the 
shell model Monte Carlo (SMMC) approach \cite{Dean.98}. The advantage of this approach 
is that it treats nuclear temperature exactly, and can even include larger model spaces. 
There are limitations, however, in applying the SMMC to odd-A and odd-odd nuclei at 
low temperatures. In addition, the SMMC yields only an averaged GT strength distribution, 
whereas the diagonalization shell-model approach allows for detailed spectroscopy.

At higher densities and core temperatures $T \approx 1$ MeV, the excitation energy of a 
nucleus with mass number $A \approx 80$ is much larger than the energy gap between 
the $pf$- and $sdg$-shells. Weak-interaction rates for nuclei beyond the $pf$-shell cannot 
yet be systematically evaluated with large-scale diagonalization shell-model calculations, 
because of huge configuration spaces and the lack  of a reliable effective interaction 
in this mass region. Thus in Ref.~\cite{LKD.01} a hybrid model has been introduced in 
which the nucleus is described as a Slater determinant with temperature-dependent 
occupation numbers, determined with shell-model Monte Carlo (SMMC) 
calculations. In the second step the electron capture rates are computed from GT$^{+}$ 
strength distributions calculated with the random-phase approximation (RPA) built on top 
of the temperature-dependent Slater determinant. The SMMC/RPA hybrid model was 
used to calculate electron capture rates on nuclei with mass numbers $A=65-112$, at 
temperatures and densities characteristic for core collapse \cite{Lan.03}. It was shown 
that these rates are so large that electron capture on nuclei dominates over capture on free 
protons, and this leads to significant changes in the hydrodynamics of core collapse and 
bounce \cite{Lan.03,Hix.03}. 

The latest theoretical and computational advances in modeling the nuclear physics input 
for astrophysical applications have highlighted the need for fully microscopic global 
predictions for the nuclear ingredients. This is especially important when considering 
neutron-rich nuclei far from the line of $\beta$-stability, for which data on ground-state 
properties and excitations are not available.
The basic advantages of the shell model is the ability to describe simultaneously all 
spectroscopic properties of low-lying states for a large domain of nuclei, and 
the use of effective interactions that can be related to two- 
and three-nucleon bare forces. On the other hand, since 
effective interactions strongly depend on the choice of active 
shells and truncation schemes, there is no universal shell-model
interaction that can be used for all nuclei. Moreover, because 
single-particle energies and a
large number of two-body matrix elements have to be adjusted to data, 
extrapolations to nuclei far from stability are not expected to be very
reliable. Medium-heavy and heavy nuclei with very large valence spaces 
require calculations with matrix dimensions that are far beyond the limits 
of current shell model variants. Properties of heavy nuclei with a large number of 
active valence nucleons are therefore best described in the framework of nuclear energy 
density functionals (NEDF). At present NEDF's provide 
the most complete description of ground-state properties and 
collective excitations over the whole nuclide chart \cite{BHR.03,VALR.05}. At the level 
of practical applications the NEDF framework is realized in terms of self-consistent 
mean-field models. With a small set of universal parameters 
adjusted to data, this approach has achieved a high level 
of accuracy in the description of structure properties over the whole 
chart of nuclides, from relatively light systems to superheavy nuclei, and from 
the valley of $\beta$-stability to the particle drip-lines.

In this work we introduce a fully self-consistent microscopic framework for calculation of 
weak-interaction rates at finite temperature, based on Skyrme functionals. The 
single-nucleon basis and the corresponding thermal occupation factors of the initial nuclear 
state are determined in the finite-temperature Skyrme Hartree-Fock model, and 
charge-exchange transitions to excited states are computed using the  
finite-temperature RPA.  Effective interactions are implemented self-consistently, i.e. 
both the finite-temperature single-nucleon Hartree-Fock equations and the 
matrix equations of RPA are based on the same Skyrme energy density functional. 
The advantage of this approach over shell-model calculations or hybrid models, is 
that a particular finite-temperature Hartree-Fock plus RPA model, i.e. completely 
determined by the choice of a Skyrme functional, can be extended over arbitrary 
mass regions of the nuclide chart, without additional assumptions or adjustment 
of parameters, as for instance single-particle energies, to transitions within 
specific shells. In a simple RPA, of course, correlations are described only  
on the one-particle -- one-hole level, and therefore one cannot expect the 
model to reproduce the details of the fragmentation of GT strength distributions. 
This can only be accomplished in the shell-model approach which includes 
higher-order correlations. In general, however, the RPA reproduces the 
centroid of strength distributions and the total GT strength. For electron capture, 
particularly, the RPA is an appropriate tool for the evaluation of cross sections 
for capture on nuclei in conditions where the electron chemical potentials 
are larger than the characteristic nuclear Q-values \cite{Jan.07}. 

Rather than evaluating and tabulating weak-interaction rates for hundreds of nuclei 
already at this stage, 
in the present work we perform illustrative calculations of electron-capture cross sections for 
selected nuclei in the iron mass group and for neutron-rich Ge isotopes, and compare  
results with those obtained with the SMMC approach \cite{Dean.98} and 
the hybrid SMMC/RPA model \cite{LKD.01}, respectively. Calculations are performed 
for a representative set of Skyrme functionals, and this provides an estimate of the 
range of theoretical uncertainty inherent in the Skyrme Hartree-Fock plus RPA approach. 

The framework of finite-temperature Skyrme Hartree-Fock plus charge-exchange 
RPA, and the formalism for calculating cross sections for electron capture, are introduced in 
Section \ref{SHFplusRPA}. Electron capture on iron group nuclei ($ A \approx 45-65$) is 
considered in Section \ref{SecFe}, and cross sections for electron capture on neutron-rich Ge
isotopes are evaluated in Section \ref{SecGe}. Section \ref{secV} 
summarizes the results of the present investigations and 
ends with an outlook for future studies.

\section{\label{SHFplusRPA} Calculation of electron capture cross sections with 
finite-temperature Skyrme Hartree-Fock plus RPA}
 
\subsection{\label{FTRPA} Charge-exchange RPA at finite temperature }

Throughout pre-supernova evolution, electron capture on nuclei proceeds 
at finite temperature. In addition to capture on $pf$-shell nuclei, this 
process also takes place on neutron-rich nuclei with protons in the 
$pf$-shell and neutron number $N > 40$. Finite temperature effects 
and correlations unblock Gamow-Teller transitions that are forbidden 
at zero temperature. 
 
In the present analysis we employ the fully self-consistent finite temperature 
charge-exchange random phase approximation (FTRPA), 
formulated in the single-nucleon basis of the Skyrme Hartree-Fock 
model at finite temperature (FTSHF). Effective interactions 
are implemented self-consistently, i.e. both the FTSHF equations and the 
matrix equations of FTRPA are based on the same Skyrme 
energy density functional. For a description of open-shell nuclei it is also 
necessary to include a consistent treatment
of pairing correlations like, for instance, in the finite temperature 
HFB+QRPA framework~\cite{Som.83,kha04}.
However, in nuclei the phase transition from a superfluid to normal
state occurs at temperatures $T \approx 0.5-1$ MeV~\cite{God1.81,God2.81, kha07}, 
whereas for temperatures above $T \approx 4$ MeV
contributions from states in the continuum become
large, and additional subtraction schemes have to be implemented to remove
the contributions of the external nucleon gas~\cite{Bon.84}. In this work 
we consider a range of temperatures relevant for the stellar electron 
capture process: $T=0.5-1.5$ MeV \cite{LKD.01}, for which 
the FTSHF plus FTRPA should provide an accurate description of the 
Gamow-Teller and forbidden transitions.

The finite temperature HF framework  \cite{Clo.67,Bon.84} has been successfully 
used in nuclear structure calculations for many years.  In the case of a Skyrme
functional the finite temperature HF equations have the same form as at zero 
temperature, but the density reads
\begin{equation}
\rho\left ( \vec r \right ) = \sum_{\alpha} f_{\alpha} \phi_{\alpha}^*(\vec r) \phi_{\alpha}(\vec r) \;,
\label{dens}
\end{equation}
where, in addition to bilinear products of single-nucleon HF wave functions $\phi_{\alpha}$, 
the contribution of each single-nucleon state is determined by the Fermi-Dirac
(FD) occupation factors 
\begin{equation}
f_{\alpha} = {1\over 1+e^{(1/kT)(\epsilon_{\alpha}-\mu)}} \; .
\end{equation}
$\epsilon_{\alpha}$ are the single-nucleon energies, and
the chemical potential $\mu$ is determined by the conservation of the 
number of nucleons $\sum_{\alpha} f_{\alpha}=A$. In this sense
the implementation of the finite temperature formalism is very similar to the 
treatment of pairing correlations in the BCS framework, 
with Fermi-Dirac factors replacing the occupancies $v_{\alpha}^2$. 
The Fermi-Dirac factors determine contributions from individual 
single-nucleon states to other types of densities as well, e.g. the kinetic energy 
density and the spin-orbit density. 

A detailed derivation of the (Q)RPA formalism at finite temperature can be found in  
Refs.~\cite{Som.83,Vau.84,Bes.84,Sag.84}, and the first self-consistent extension 
to open-shell nuclei (finite temperature QRPA) has been recently reported in Ref.~\cite{kha04}. 
Finite-temperature linear response theory and RPA have been
successfully used in numerous studies of giant resonances and decay of hot nuclei 
\cite{Som.83,Bon.84,Bes.84,Sag.84,Vau.84,Bor.86,Ala.90,Cho.90,Rei.01,ER.93,Lac.98,Sto.04}.
The FTRPA represents the small amplitude limit of the time-dependent mean-field theory 
at finite temperature. Starting from the response of a time-dependent density matrix $\rho(t)$ 
to a harmonic external field $f(t)$ \cite{RingSchuck}, the equation of motion for the density 
operator reads
\begin{equation}
\label{EOM}
 i\partial_t \hat{\rho} = [ \hat{h}[ \hat{\rho}]+ \hat{f}(t), \hat{ \rho} ] \; .
\end{equation}
In the small amplitude limit the density matrix is expanded to linear order
\begin{equation}
 \hat{\rho} (t)= \hat{\rho}^{0}+ \delta \hat{\rho}(t) \;,
 \end{equation}
where 
\begin{equation}
\delta\hat{\rho}(t) = \delta\hat{\rho}^{(+)}e^{-i\omega t} + \delta\hat{\rho}^{(-)}e^{+i\omega t} \;,
\end{equation}
and $\hat{\rho}^{0}$ denotes the stationary ground-state density
\begin{equation}
\rho^{0}_{\alpha\beta}= \delta_{\alpha\beta}f_{\alpha} = \delta_{\alpha\beta}
 [1+e^{(1/kT) ({\epsilon_{\alpha} - \mu})}]^{-1} \; ,
 \end{equation}
and includes the thermal occupation factors of single-particle states $f_k$. For the analogous 
expansion of the Hamiltonian operator $\hat{h}(t)=\hat{h}^{(0)} + \delta \hat{h}(t)$, the
linearized equation of motion reads
\begin{equation}
i\hbar\ \partial_t\delta\hat{\rho} = \left[ \hat{h}^{(0)}, \delta\hat{\rho} \right] +
 \left[ {\delta\hat{h}\over\delta\rho} \delta\hat{\rho}, \hat{\rho}^{(0)} \right] \; .
\end{equation}
Taking the matrix elements of this equation between the states
$\langle \alpha\beta^{-1}\vert$ and $\vert 0\rangle$, we obtain
\begin{eqnarray} 
\hbar\omega \delta\rho^{(+)}_{\alpha\beta} & = & \left( \epsilon_\alpha
- \epsilon_\beta \right) \delta\rho^{(+)}_{\alpha\beta} 
+ \sum_{\gamma\delta} (f_\delta-f_\gamma) v_{\alpha\bar\delta\bar\beta\gamma} 
\delta\rho^{(+)}_{\gamma\delta} + 
(f_\gamma-f_\delta) v_{\alpha\gamma\bar\beta\bar\delta} 
\delta\rho^{(-)}_{\gamma\delta}\;, \nonumber \\
-\hbar\omega \delta\rho^{(-)}_{\alpha\beta} & = & \left( \epsilon_\alpha
- \epsilon_\beta \right) \delta\rho^{(-)}_{\alpha\beta} 
+ \sum_{\gamma\delta} (f_\gamma-f_\delta) v_{\alpha\gamma\bar\beta\bar\delta} 
\delta\rho^{(+)}_{\gamma\delta} + 
(f_\delta-f_\gamma) v_{\alpha\bar\delta\bar\beta\gamma}
\delta\rho^{(-)}_{\gamma\delta} \;,  
\label{RPA1}
\end{eqnarray} 
for terms multiplying the factors $e^{-i\omega t}$ and $e^{+i\omega t}$, respectively.  
\begin{equation}  
\delta h^{(+)}_{\alpha\beta} = \sum_{\gamma\delta} 
(f_\delta-f_\gamma) 
v_{\alpha\bar\delta\bar\beta\gamma}\delta\rho^{(+)}_{\gamma\delta} +
(f_\gamma-f_\delta) 
v_{\alpha\gamma\bar\beta\bar\delta}\delta\rho^{(-)}_{\gamma\delta}
\end{equation}
and analogously for $\delta h^{(-)}_{\alpha\beta}$. 
These relations are consistent with the definition of the HF
mean field at finite temperature, and express the
fact that thermal occupancies determine the way 
density fluctuations affect the mean field. Equations (\ref{RPA1}) are consistent 
with those defined in Ref. \cite{Vau.84}. If the first (second) equation 
of the set (\ref{RPA1}) is multiplied by $f_\beta-f_\alpha$ ($f_\alpha-f_\beta$), 
with the definition $F_{\alpha\beta} = \delta\rho^{(+)}_{\alpha\beta}$ and 
$F_{\beta\alpha} = \delta\rho^{(-)}_{\alpha\beta}$, then equations 
(\ref{RPA1}) take the form as in Ref. \cite{Bes.84}. 
The set of equations (\ref{RPA1}) is also consistent with the
derivation of finite temperature QRPA in Ref. \cite{Som.83}, 
in the limit of vanishing pairing correlations.
However, in \cite{Som.83}  a different definition of
the RPA amplitudes is introduced, which involves quantities 
like $\sqrt{f_\alpha-f_\beta}$. This requires special care
in the proton-neutron case because no simple condition can be
imposed which would guarantee that the quantity under the square 
root is positive. 

The finite temperature forward- and backward-going amplitudes can
be related to the corresponding zero-temperature amplitudes ($X$ and $Y$) 
through the following relations
\begin{equation}
\delta\rho^{(+)}_{\alpha\beta} = X_{\alpha\beta}f_\beta(1-f_\alpha)
+ Y_{\beta\alpha}f_\alpha(1-f_\beta) \;,
\end{equation} 
and
\begin{equation}
\delta\rho^{(-)}_{\alpha\beta} = Y_{\alpha\beta}f_\beta(1-f_\alpha)
+ X_{\beta\alpha}f_\alpha(1-f_\beta) \; .
\end{equation} 
The charge-exchange RPA matrices are composed of matrix elements of 
the residual interaction $v$, as well as certain combinations of 
thermal occupation factors $f_k$.  Because of finite temperature, the 
configuration space includes particle-hole ($ph$), 
particle-particle ($pp$), and hole-hole ($hh$) proton-neutron pairs.
The residual interaction is derived from a Skyrme energy density functional,
and single-particle occupation factors at finite temperature are included
in a consistent way both in the FTSHF and FTRPA. The same
interaction is used both in the FTSHF equations that determine the single-nucleon
basis, and in the matrix equations of the FTRPA. The full set of
FTRPA equations is solved by diagonalization. The result are excitation 
energies, and the corresponding forward- and
backward-going amplitudes that are used to evaluate the transition strength for 
a given multipole operator.
\begin{figure}
\centerline{
\includegraphics[scale=0.5,angle=0]{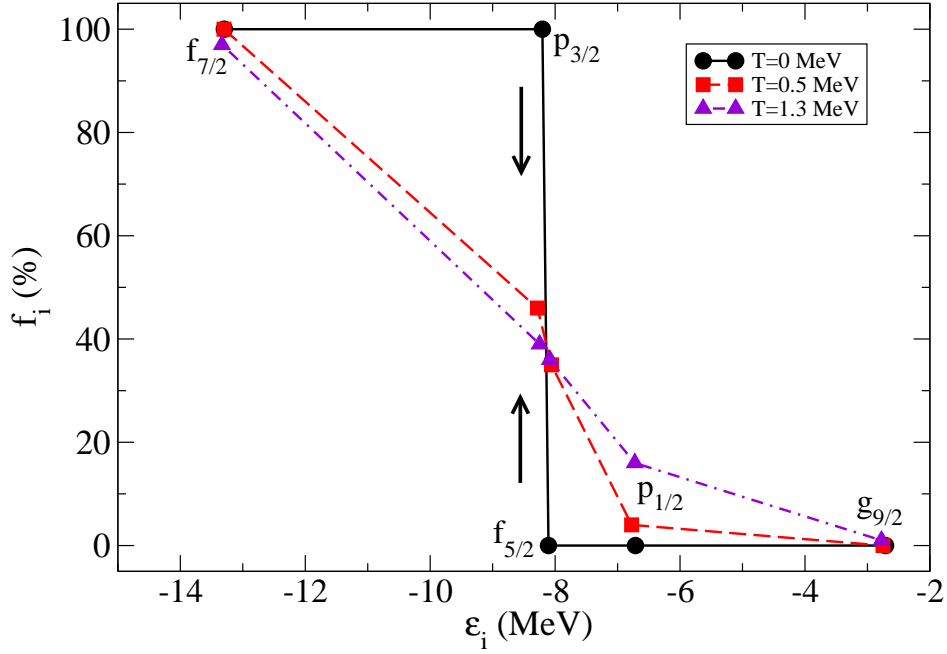}
}
\caption{(Color online) Occupation percentage of the proton orbitals 
f$_{7/2}$, p$_{3/2}$, f$_{5/2}$, p$_{1/2}$ and g$_{9/2}$ 
in $^{74}$Ge, calculated in the finite-temperature Skyrme HF model 
with the SGII interaction, at zero temperature, $T=0.5$ MeV and 
$T=1.3$ MeV. }
\label{protsub}
\end{figure}
\begin{figure}
\centerline{
\includegraphics[scale=0.5,angle=0]{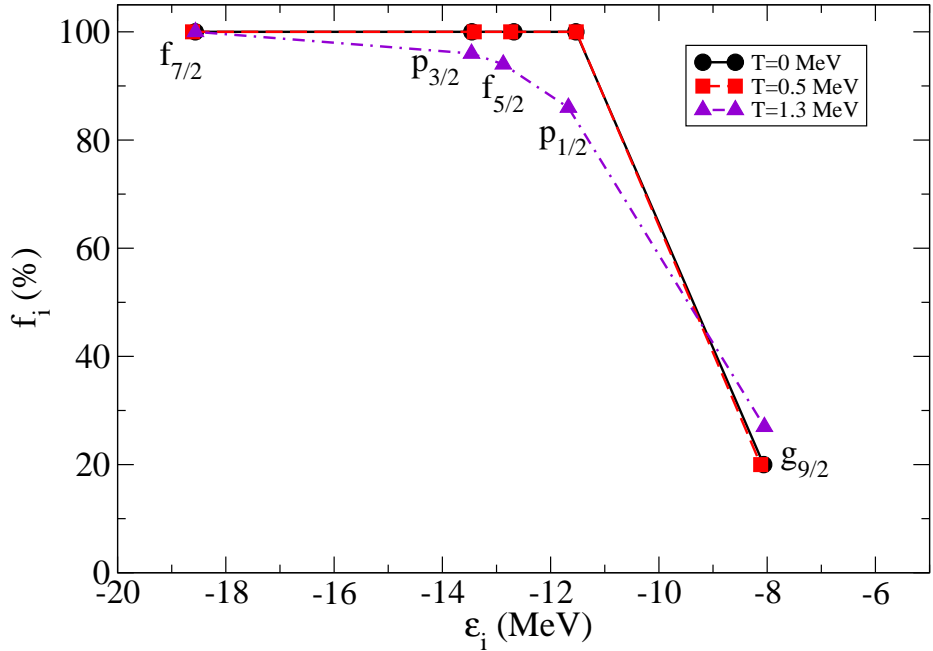}
}
\caption{(Color online) Occupation percentage of the neutron orbitals 
f$_{7/2}$, p$_{3/2}$, f$_{5/2}$, p$_{1/2}$ and g$_{9/2}$ 
in $^{74}$Ge, calculated in the finite-temperature Skyrme HF model 
with the SGII interaction, at zero temperature, $T=0.5$ MeV and 
$T=1.3$ MeV.}
\label{neutsub}
\end{figure}
As an illustrative example, in Figs.~\ref{protsub} and \ref{neutsub} we display 
the temperature dependence of the occupations of the proton and neutron 
orbitals f$_{7/2}$, p$_{3/2}$, f$_{5/2}$, p$_{1/2}$ and g$_{9/2}$ in 
$^{74}$Ge, calculated in the finite-temperature Skyrme HF model. 
The self-consistent calculation with the SGII effective interaction \cite{gia81} corresponds to 
zero temperature, $T=0.5$ MeV and $T=1.3$ MeV. At zero temperature the proton 
 orbitals f$_{7/2}$ and p$_{3/2}$ are fully occupied, whereas f$_{5/2}$, p$_{1/2}$ and 
 g$_{9/2}$ are empty. By increasing the temperature protons are mostly 
 promoted from the p$_{3/2}$ into the f$_{5/2}$ orbital and, to a lesser extent, 
into the p$_{1/2}$ orbital. Correspondingly the occupation percentage of 
p$_{3/2}$ is reduced, whereas those of the f$_{5/2}$ and p$_{1/2}$ orbitals 
are enhanced. In the temperature intervals considered here, the occupations 
of the f$_{7/2}$ and  g$_{9/2}$ orbitals do not change significantly. At zero 
temperature neutrons fully occupy the f$_{7/2}$, p$_{3/2}$, f$_{5/2}$, and p$_{1/2}$ 
orbitals, and there are two neutrons in the g$_{9/2}$ orbital. The main effect of 
increasing temperature is to promote neutrons from p$_{3/2}$, f$_{5/2}$, and p$_{1/2}$ 
into g$_{9/2}$. However, note that for protons already at $T=0.5$ MeV the calculation 
predicts a pronounced effect on the occupation of orbitals close to the Fermi surface, 
whereas the occupation of neutron orbitals is significantly modified only at the 
higher temperature of $T=1.3$ MeV. The results shown in Figs.~\ref{protsub} and 
\ref{neutsub} can be compared with thermal occupation numbers obtained
from canonical shell model Monte Carlo (SMMC) calculated in Ref.~\cite{LKD.01}, 
with Woods-Saxon single-particle energies and a pairing plus quadrupole 
residual interaction. The temperature dependence of occupation numbers 
predicted by the two models is similar, with the SMMC results showing a more 
pronounced effect on occupation numbers already at $T=0.5$ MeV, 
especially for neutron orbitals. This can be attributed to a smaller energy 
gap between the p$_{1/2}$ and g$_{9/2}$ single-neutron orbitals used in the 
SMMC calculation, and to additional correlations in the ground state that are 
not taken into account in the simple Skyrme HF model.
\begin{figure}
\centerline{
\includegraphics[scale=0.5,angle=0]{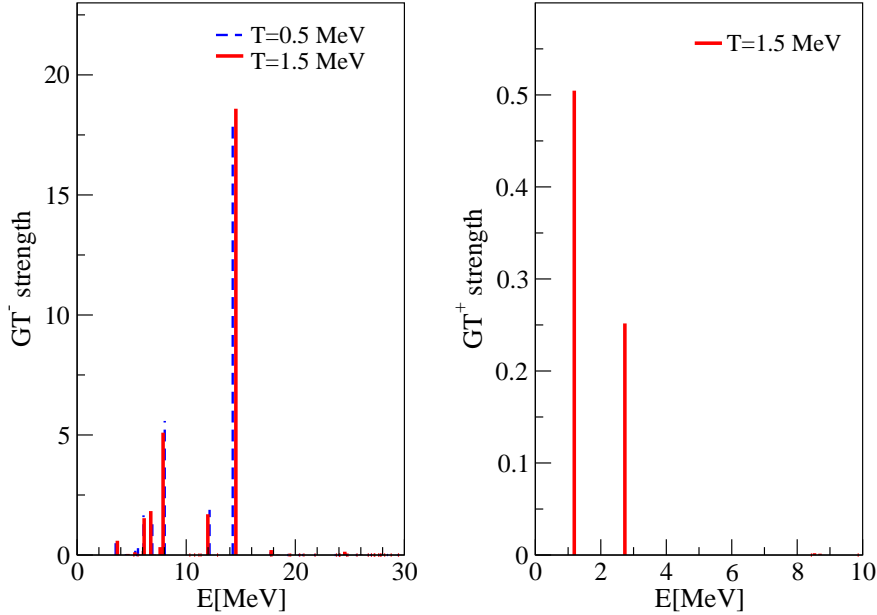}
}
\caption{(Color online) Temperature dependence of the Gamow-Teller GT$^{-}$ 
and GT$^{+}$
strength distributions in $^{74}$Ge, calculated with the finite-temperature 
proton-neutron RPA model based on the Skyrme SGII interaction.}
\label{gtmin}
\end{figure}

In Fig. \ref{gtmin} we display the temperature dependence of the corresponding 
Gamow-Teller GT$^{-}$ and GT$^{+}$ strength distributions in $^{74}$Ge, calculated with the 
finite-temperature proton-neutron RPA model based on the Skyrme SGII interaction.
In the GT$^{-}$ direction a neutron is changed into a proton, as in $\beta$-decay.
At low temperature the GT$^{-}$ mode corresponds to a coherent superposition of 
$J^{\pi}=1^{+}$ charge-exchange proton-particle -- neutron-hole transitions. 
The GT operator reads
\begin{equation}
T_{GT}^{\pm}=\sum_{i=1}^{A}\bm{\sigma}\tau_{\pm} \; .
\label{gtopera}
\end{equation}
In addition to the high-energy GT$^-$  
resonance at $\approx 14$ MeV, a collective superposition of 
direct spin-flip ($j = l + \frac{1}{2}$ $\rightarrow$ $j = l - \frac{1}{2}$) 
transitions, the response function displays a concentration of
strength in the low-energy tail. The transitions in the low-energy
region correspond to core-polarization 
($j = l \pm \frac{1}{2}$ $\rightarrow$ $j = l \pm \frac{1}{2}$), 
and back spin-flip 
($j = l - \frac{1}{2}$ $\rightarrow$ $j = l + \frac{1}{2}$)
neutron-hole -- proton-particle excitations. The GT$^{-}$ transitions are allowed 
at $T=0$, and the distribution displays only a weak temperature dependence. 
On the other hand, for $^{74}$Ge the GT$^{+}$ transitions, in which a proton is 
changed into a neutron, are forbidden at zero temperature because the 
relevant neutron orbitals are fully occupied. Note that additional ground-state 
correlations, e.g. pairing,  are not taken into account in the Skyrme Hartree-Fock plus 
RPA model. Even at  $T=0.5$ MeV the occupation factors of neutron orbitals remain 
unchanged (cf. Fig.~\ref{neutsub}) and the calculation does not predict any 
low-energy GT$^{+}$ transition. Only at higher temperatures above 
$T=1.0$ MeV the low-energy GT$^{+}$ transitions are thermally unblocked.
We have verified that at each temperature the 
Gamow-Teller strength distribution satisfies the Ikeda sum rule \cite{Ike.63}
\begin{equation}
\left(  S_{GT}^{-}-S_{GT}^{+}\right)  =3(N-Z) \;,
\label{gtsrule}
\end{equation}
where $S_{GT}^{\pm}$ denotes the total sum of Gamow-Teller
strength for the GT$^{\pm}$ transitions. 

\subsection{\label{cross_section} Cross section for electron capture}

The electron capture on a nucleus $(Z, N)$ 
\begin{equation}
e^{-} + ({Z},{N}) \longrightarrow  \nu_{e} + ({Z-1},{N+1})^{*} \; ,
\end{equation}
presents a simple semi-leptonic reaction that proceeds via the charged 
current of the weak interaction. The theoretical analysis of these processes
necessitates the description of the weak interaction between leptons and 
nucleons, as well as the wave functions of the initial and final nuclear states.
Detailed expressions for the reaction rates and the transition matrix elements 
can be found in Refs.~\cite{O'Connell1972,Walecka75,Walecka2004}.
The electron-nucleus reaction cross section for a transition between 
the states $| i \rangle$ and $| f \rangle$ reads
\begin{equation}
\frac{d\sigma}{d\Omega} = \frac{V E_{\nu}^{2}}{(2\pi)^2} \sum_{\rm  
lepton\, spins} \frac{1}{2J_{i} + 1}\sum_{M_{i}}\sum_{M_{f}} \left|  
\langle f\right| \hat{H}_{W} \left| i \rangle \right|^{2}\; ,
\end{equation}
where $V$ denotes the quantization volume,
and  $E_{\nu}$ is the energy of the outgoing electron neutrino.  
The Hamiltonian $ \hat{H}_{W} $ of the weak interaction 
is expressed in the standard current-current form, i.e. in terms of the 
nucleon $\mathcal{J}_{ \lambda }(\bm{x})$ and lepton $j_{ \lambda }(\bm{x})$ 
currents
\begin{equation}
\hat{H}_{W}=- \frac{G }{ \sqrt{2} }  \int d^3{x}~
\mathcal{J}_{ \lambda }(\bm{x})j^{ \lambda }(\bm{x}) \;,
\end{equation}
where $G$ is the weak coupling constant, and the resulting 
transition matrix element reads
\begin{equation}
\langle f | \hat{H}_{W}| i \rangle = -  \frac{ G}{\sqrt{2}}l_{ \lambda }
  \int  \frac{d^3{x}  }{1/\sqrt{V}}~e^{- i \bm{q} \cdot \bm{x}}
\langle f | \mathcal{J}^{ \lambda }(\bm{x}) | i \rangle \; ,
\end{equation}
where the four-momentum transfer is $q \equiv (q_0, \bm{q})$, and 
the multipole expansion of the leptonic matrix element 
$l_{ \lambda }  e^{- i \bm{q} \cdot \bm{x}}$ determines 
the operator structure for the nuclear transition matrix elements 
\cite{O'Connell1972,Walecka75,Walecka2004}. The expression for  
the electron capture cross sections is given by
\begin{eqnarray}
  & &\frac{d \sigma}{ d \Omega }
  =  \frac{ G_F^2cos^2 \theta_c }{2\pi}
  \frac{F(Z,E_e)}{(2J_i+1)} \nonumber \\
& &\times  \Bigg\{ \sum_{J \geq 1} \mathcal{W}(E_{\nu})
  \Big\{
  {\left(
1-(\hat{\bm{\nu}} \cdot \hat{\bm{q}})(\bm{\beta} \cdot \hat{\bm{q}})
  \right )}
  \left[ \vert \langle J_f || \hat{\mathcal{T}}_J^{MAG} || J_i  
\rangle \vert^2
+ \vert \langle J_f || \hat{\mathcal{T}}_J^{EL} || J_i \rangle  
\vert^2 \ \right ] \nonumber \\
& & -  2\hat{\bm{q}} \cdot (\hat{\bm{\nu}} -  \bm{\beta} )
Re\langle J_f || \hat{\mathcal{T}}_J^{MAG} || J_i \rangle
\langle J_f || \hat{\mathcal{T}}_J^{EL}|| J_i \rangle^{*} \Big\}  
\nonumber \\
& & +\sum_{J \geq 0 } \mathcal{W}(E_e,E_{\nu}) \Big\{
(1-\hat{\bm{\nu}} \cdot  \bm{\beta} + 2(\hat{\bm{\nu}} \cdot \hat{\bm 
{q}})(\bm{\beta} \cdot \hat{\bm{q}})
\langle J_f || \hat{\mathcal{L}}_J || J_i \rangle | ^2
+ (1+ \hat{\bm{\nu}} \cdot  \bm{\beta}) \langle J_f || \hat{\mathcal 
{M}}_J || J_i \rangle | ^2 \nonumber \\
& & - 2 \hat{\bm{q}} (\hat{\bm{\nu}} +  \bm{\beta}) Re \langle J_f ||  
\hat{\mathcal{L}}_J || J_i \rangle
\langle J_f || \hat{\mathcal{M}}_J|| J_i \rangle^{*} \Big\} \Bigg\} \;,
\label{ec_rate}
\end{eqnarray}
where the momentum transfer $\bm{q}=\bm{\nu}-\bm{k}$ is defined as the
difference between neutrino and
electron momenta, $\hat{\bm{q}}$ and $\hat{\bm{\nu}}$
are the corresponding unit vectors, and $\bm{\beta} = \bm{k}/ 
E_e$. The energies of the incoming electron and outgoing neutrino are denoted by
$E_e$ and $E_{\nu}$, respectively. The Fermi function $F(Z,E_e)$ corrects the 
cross section for the distortion of the electron wave function by the Coulomb field of 
the nucleus \cite{Kol.03}. 
\begin{eqnarray}
  \mathcal{W}(E_{\nu}) = \frac{E_{\nu}^2}{(1+E_{\nu}/M_T)}\; ,
\end{eqnarray}
with the phase-space factor $(1+E_{\nu}/M_{T})^{-1}$ accounting
for the nuclear recoil, and $M_T$ is the mass of the target nucleus.
The nuclear transition matrix elements between the initial state
$|J_i \rangle$ and final state $|J_f \rangle$, correspond to the
charge $\hat{\mathcal{M}}_J$, longitudinal $\hat{\mathcal{L}}_J$,
transverse electric $ \hat{\mathcal{T}}_J^{EL}$, and
transverse magnetic $\hat{\mathcal{T}}_J^{MAG}$ multipole operators:
\begin{itemize}
\item the Coulomb operator
\begin{equation}
\hat{\mathcal{M}}_{JM}(\bm{x})= F_1^V M_J^M(\bm{x})
- i \frac{ \kappa  }{m_N} 
\left[  F_A  \Omega _J^M(\bm{x})
+ \frac{ 1}{ 2}  (F_A- m_{e} F_P)  \Sigma''^M_J(\bm{x})   \right ]  \;,
\label{coulombop}
\end{equation}
\item the longitudinal operator
\begin{equation}
\hat{\mathcal{L}}_{JM}(\bm{x})=  \frac{q_0}{ \kappa }F_1^V M_J^M(\bm{x})
+i F_A \Sigma ''^M_J(\bm{x}) \;,
\label{longitudinalop}
\end{equation}
\item the transverse electric operator
\begin{equation}
\hat{\mathcal{T}}_{JM}^{el}(\bm{x})=  \frac{\kappa }{ m_N}
 \left[ F^V_1  {\Delta'}_{J}^{M}(\bm{x})+ \frac{1 }{2 } \mu^V
 \Sigma_J^M(\bm{x})     \right ]
+i F_A  {\Sigma'}_J^M (\bm{x}) \;,
\label{transverseelop}
\end{equation}
\item and the transverse magnetic operator
\begin{equation}
\hat{\mathcal{T}}_{JM}^{mag}(\bm{x}) = -i  \frac{ \kappa  }{m_N }
 \left[ F_1^V  \Delta_J^M(\bm{x})- \frac{1 }{2 }  \mu^V  {\Sigma'}_J^M(\bm{x})     \right ]
+F_A  \Sigma_J^M(\bm{x}) \; ,
\label{transversemagop}
\end{equation}
\end{itemize}
where all the form factors are functions of $q^2$, and
$\kappa = \left| \bm{q} \right|$.
The operators $M$, $\Omega$, $\Delta$, $\Delta'$, $\Sigma$, $\Sigma'$, and $\Sigma''$ are 
expressed in terms of  spherical Bessel functions,
spherical harmonics, and vector spherical harmonics~\cite{O'Connell1972}.
By assuming conserved vector current (CVC), the standard 
set of form factors reads \cite{Kuramoto1990}: 
\begin{equation}
F_{1}^{V} (q^{2}) = \left[ 1 + \left( \frac{q}{840\,\textrm{MeV}}  
\right)^{2} \right]^{-2},
\label{ff1}
\end{equation}
\begin{equation}
\mu^{V} (q^{2}) = 4.706 \left[ 1 + \left( \frac{q}{840\,\textrm{MeV}}  
\right)^{2} \right]^{-2},
\label{ff2}
\end{equation}
\begin{equation} \label{eq:fa}
F_{A} (q^{2}) = -1.262 \left[ 1 + \left( \frac{q}{1032\,\textrm{MeV}}  
\right)^{2} \right]^{-2},
\label{ff3}
\end{equation}
\begin{equation} \label{eq:fp}
F_{P} (q^{2}) = \frac{2 m_{N} F_{A}(q^{2})}{q^{2} + m_{\pi}^{2}} \; .
\label{ff4}
\end{equation}
\bigskip

The cross sections for electron capture are evaluated from Eq.~(\ref{ec_rate}), 
with transition matrix elements between the initial and final states  
determined in a self-consistent microscopic framework based on 
the (finite temperature) Skyrme HF model for the  
nuclear ground state, and excited states are calculated using the 
corresponding (finite temperature) RPA. For each transition operator $\hat{O}_J$ 
the matrix elements between the initial state of the even-even $(Z, N)$ target nucleus
and the final state in the corresponding $(Z-1,N+1)$ nucleus are  
expressed in terms of single-particle matrix elements
between the single-nucleon states, and the corresponding (finite temperature) RPA
amplitudes  $\delta \rho^{(+)J}_{\alpha\beta}$ and
$\delta \rho^{(-)J}_{\alpha\beta}$ (cf. Sec.~\ref{FTRPA}):
\begin{equation}
  \langle J_f || \hat{O}_J || J_i \rangle =
  \sum_{\alpha\beta}
  \langle \alpha || \hat{O}_J || \beta \rangle
\left( \delta \rho^{(+)J}_{\alpha\beta}   -  \delta\rho^{(-)J}_{\alpha 
\beta} \right).
\label{redtrans}
\end{equation}
The energy of the outgoing neutrino is determined by the 
conservation relation:
\begin{equation}
E_{\nu} = E_e - Q + E_i - E_f \; ,
\end{equation}
which includes the difference between the final and initial nuclear states.   
The $Q$-value plays a particularly important role in the calculation of electron 
capture rates. Namely, the energy that is available to excite states in the daughter 
nucleus depends on whether electron capture on a specific target nucleus releases 
energy ($Q < 0$), or requires an additional external input ($Q > 0$). 
In the present calculation the $Q$-value is determined from the experimental
masses \cite{AWT}: $Q=M_f - M_i$, where $M_{i,f}$ are the masses of the parent and 
daughter nuclei, respectively.

The nuclei that will be considered in this work contribute to stellar electron 
capture rates in the temperature interval $T \approx 0.5-1.5$ MeV~\cite{LKD.01}.  
The expression for the total cross section for electron capture on a nucleus 
$(Z, N)$ at temperature $T$ reads
\begin{equation}
\sigma (E_e,T) = \frac{G^2}{2 \pi} \sum_{i}
F(Z,E_e) \frac{(2J_i+1) e^{-E_i/(kT)}}{G(Z,A,T)}
\sum_{f,J}
{(E_e- Q + E_i - E_f)^2} \frac{|\langle i| \hat{O}_J | f \rangle|^2}{(2J_i+1)}\;,
\end{equation}
where $\hat{O}_J$ is the generic notation for the 
charge $\hat{\mathcal{M}}_J$, longitudinal $\hat{\mathcal{L}}_J$,
transverse electric $ \hat{\mathcal{T}}_J^{EL}$, and
transverse magnetic $\hat{\mathcal{T}}_J^{MAG}$ multipole operators. 
The sum over initial states includes a thermal
average of levels, with the corresponding partition function $G(Z,A,T)$. 
The finite temperature induces the thermal population of excited states 
in the parent nucleus. Each of these states $| i \rangle$ is connected 
by the multipole operators to many levels $| f \rangle$ in the daughter nucleus. 
The calculation of all possible transitions is computationally prohibitive, and 
therefore the evaluation of the total cross section for electron capture is 
usually simplified \cite{FFN,Aufderheide94,LM.00,LKD.01} by adopting the 
Brink hypothesis, i.e. by assuming that the
strength distribution of the multipole operators in the daughter nucleus
is the same for all
initial states and shifted by the excitation energy of the initial
state.  By using this approximation, the sum over final states becomes
independent of the initial state and the sum over the Boltzmann weights
cancels the partition function. The Brink hypothesis is a valid approximation 
when the temperature and density are high enough so that many 
states contribute, and variations in the low-energy transition strength cancel out. 
As it was done in the calculation of stellar electron capture on neutron-rich 
germanium isotopes in Ref.~\cite{LKD.01}, we apply the Brink hypothesis 
to the initial state  which represents the thermal average of many-body states in
the parent nucleus at temperature $T$. This thermally
averaged initial state is approximated by the finite temperature 
Skyrme Hartree-Fock ground-state Slater determinant with
Fermi-Dirac thermal occupation factors. With this approximation the 
final expression for the total electron capture cross section at temperature 
$T$ reads
\begin{equation}
\label{tot_sigma}
\sigma (E_e,T) = \frac{G^2}{2 \pi} F(Z,E_e)
\sum_f {(E_e-Q - \omega_f)^2} \sum_J S_J (\omega_f,T) \; ,
\end{equation}
where $\omega_f$ is the excitation energy in the daughter nucleus, and
$S_J$ is the discrete finite-temperature RPA response for the multipole operator 
$\hat{O}_J$.

\section{\label{SecFe} Electron capture on iron group nuclei}

As our first illustrative example and application of the model, we consider electron capture 
on iron group nuclei ($ A \approx 45-65$). The calculated electron capture cross sections 
and rates for nuclei in this mass range are essential for modeling the initial phase 
of stellar core-collapse and supernova 
explosion \cite{Bet.90,Aufderheide94,LM.RMP03,Jan.07}. In the presupernova 
collapse electron capture on $pf$-shell nuclei proceeds at temperatures between 
300 keV and 800 keV. Detailed calculations of stellar weak-interaction rates in the iron 
mass region have been carried out in the framework of the interacting shell 
model \cite{LM.RMP03}. Both the shell model Monte Carlo (SMMC) \cite{Dean.98}, 
and large-scale shell-model diagonalization \cite{LM.00,Mar.00}, were used to calculate 
electron capture and $\beta$-decay rates in the $ A \approx 45-65$ mass region. 
The SMMC results \cite{Dean.98} were in fact superseded in Ref.~\cite{LM.00}, by the 
weak-interaction rates obtained using large-scale shell-model diagonalization in the 
complete $pf$-shell.

In this section we calculate electron capture cross sections for selected nuclei in the 
iron mass region, and compare the results with those of Ref.~\cite{Dean.98}, where the 
SMMC was used to calculate Gamow-Teller GT$^{+}$ strength distributions (in this 
direction a proton is changed into a neutron), and these distributions were then used 
to compute the electron-capture cross sections and rates in the zero-momentum transfer 
limit, as a function of the incident energy of the electron. The SMMC calculations solve 
the full shell-model problem for the GT$^{+}$ strength distributions in the $0\hbar\omega$ 
$fp$-shell space using a realistic residual interaction. In the calculation of Ref.~\cite{Dean.98} 
the KB3 residual interaction \cite{PZ.81}  was used, and the quenching of the total GT strength 
was taken into account by renormalizing the GT transition matrix elements by the constant 
factor 0.8. In the present analysis the cross sections for electron capture are evaluated from
Eq.~(\ref{tot_sigma}), with transition matrix elements between the initial and final states  
determined in the self-consistent microscopic framework based on 
the finite temperature Skyrme HF model for the  
nuclear ground state, and excited states are calculated using the 
corresponding finite temperature charge-exchange RPA.
\begin{figure}
\centerline{
\includegraphics[scale=0.6,angle=0]{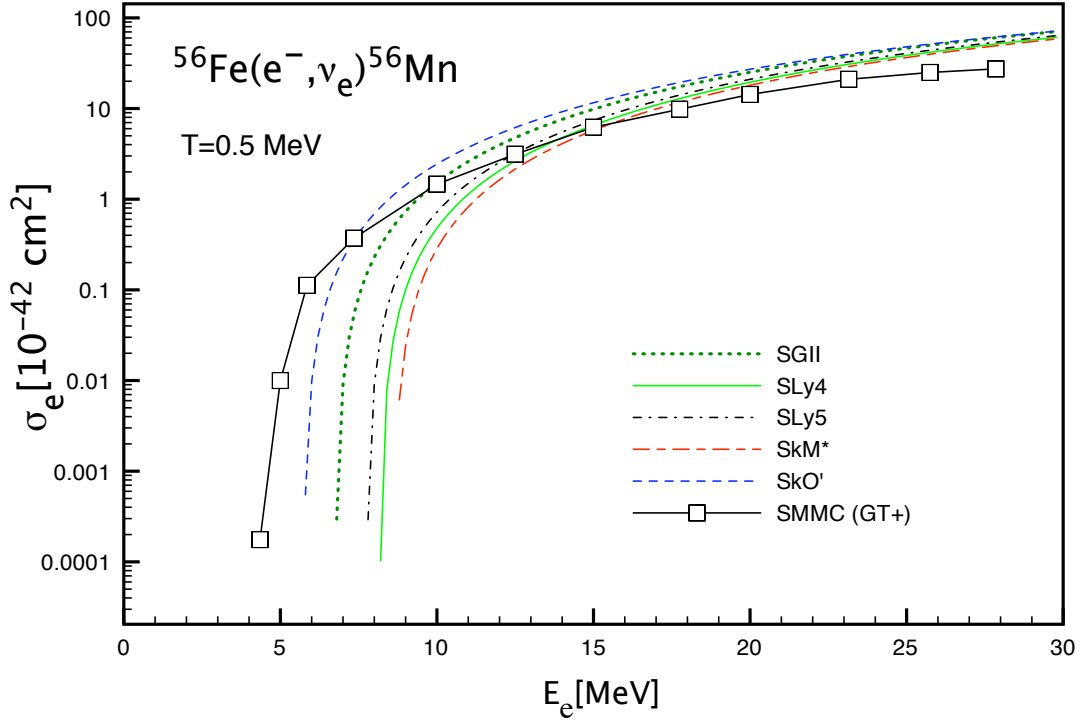}
}
\caption{(Color online) Electron capture cross sections for the reaction 
$^{56}$Fe$(e^-,\nu_e)^{56}$Mn at $T=0.5$ MeV, as functions 
of the incident electron energy $E_e$. The 
Skyrme HF+RPA results are compared with cross 
sections calculated from the SMMC Gamow-Teller strength 
distributions \cite{Dean.98}.}
\label{eccs_Fe56}
\end{figure}

In Fig.~\ref{eccs_Fe56} we display the calculated 
cross sections for the reaction $^{56}$Fe$(e^-,\nu_e)^{56}$Mn 
as functions of the incident electron energy $E_e$.
Calculations are performed at temperature T=0.5 MeV,
and the Skyrme HF+RPA results are compared with cross 
sections calculated from SMMC Gamow-Teller strength 
distributions \cite{Dean.98}. As the calculation of cross sections in 
Ref.~\cite{Dean.98} corresponds to the zero-momentum transfer limit 
and only includes the Gamow-Teller operator, for the sake of 
comparison we have also limited the sum in Eq.~(\ref{ec_rate}) to 
the $1^+$ channel, i.e.  only transitions to $1^+$ excited states are 
taken into account. At low momentum transfer, the  allowed Gamow-Teller 
transitions dominate the electron capture process on $pf$-shell nuclei. 
Note, however, that in the calculation of Dean {\em et al.} \cite{Dean.98} 
only the $0\hbar \omega$ Gamow-Teller transition strength 
is considered, rather than the total strength in the $1^+$ channel.  
The reduction of the axial-vector coupling constant from its 
free-nucleon value $g_A = 1.262$ to $g_A = 1.0$ (cf. Eq.~(\ref{ff3})), 
is equivalent to the renormalization of the GT matrix elements in 
Ref.~\cite{Dean.98}  by the constant factor 0.8. In the shell-model 
studies of  weak-interaction rates in the $ A \approx 45-65$ mass region,  
both the SMMC \cite{Dean.98} and shell-model 
diagonalization approach \cite{LM.00}, have used the KB3 residual 
interaction, which is well suited for full $0 \hbar \omega$ calculations 
throughout the lower $pf$-shell region. However, to calculate the 
weak-interaction rates in the entire mass range $ A \approx 45-65$, 
the original KB3 interaction had to be modified by including a 
number of monopole corrections in order to reproduce the GT strength 
distributions and half-lives. 

\begin{figure}
\centerline{
\includegraphics[scale=0.6,angle=0]{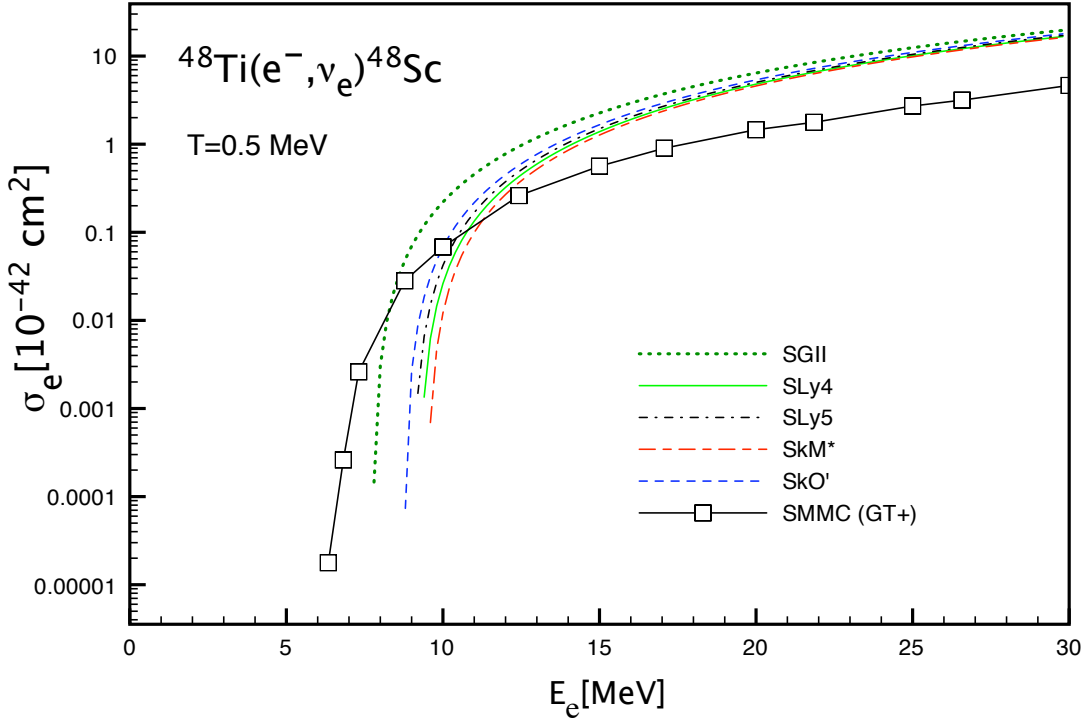}
}
\caption{(Color online) Same as in Fig.~\ref{eccs_Fe56}, but for the 
reaction $^{48}$Ti$(e^-,\nu_e)^{48}$Sc .}
\label{eccs_Ti48}
\end{figure}

\begin{figure}
\centerline{
\includegraphics[scale=0.6,angle=0]{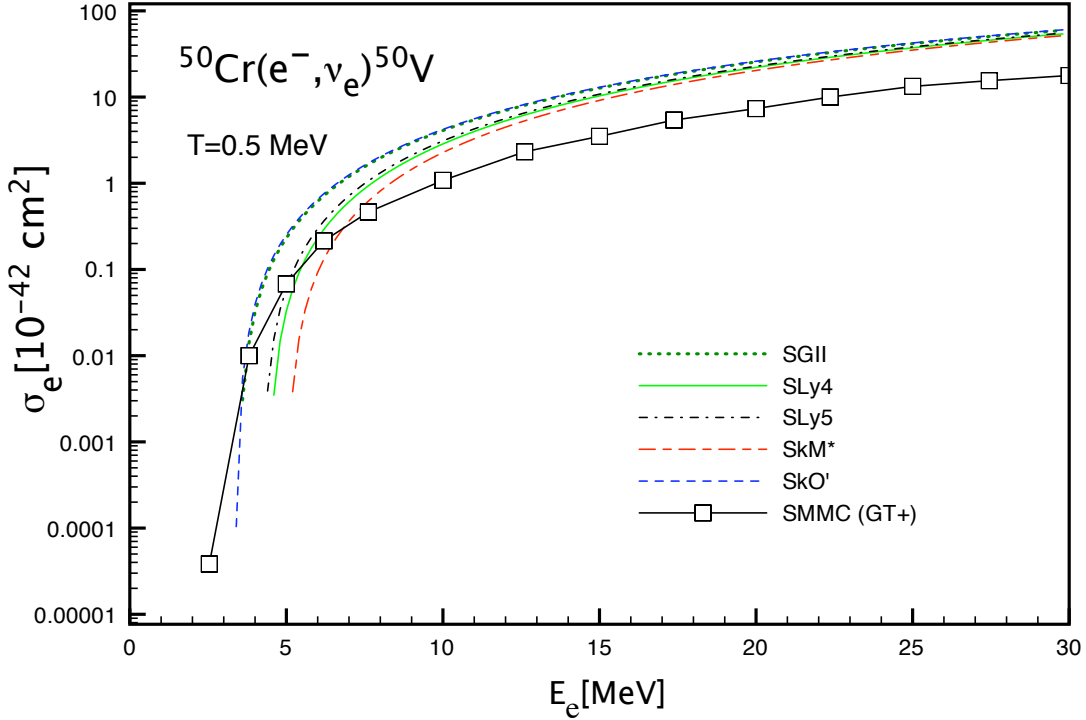}
}
\caption{(Color online) Same as in Fig.~\ref{eccs_Fe56}, 
but for the reaction $^{50}$Cr$(e^-,\nu_e)^{50}$V.}
\label{eccs_Cr50}
\end{figure}

In the present analysis the cross sections, as functions of the incident 
electron energy, are computed for a representative set of Skyrme 
functionals: SGII \cite{gia81}, SkM* \cite{bar82}, SLy4\cite{Cha.97}, 
SLy5\cite{Cha.98},  and SkO' \cite{Rei.99}. 
Over the last thirty years more than hundred different Skyrme parameterizations 
have been adjusted and analyzed, and it is often difficult to compare 
results obtained with different models, also because they include different 
subsets of terms from the most general functional. Since in this work 
we apply the microscopic approach based on Skyrme 
HF+RPA, calculations are performed using various 
Skyrme functionals. In principle this will provide an
estimate of the range of theoretical uncertainty inherent in the present 
approach. The electron-capture cross sections in Fig.~\ref{eccs_Fe56} 
exhibit a sharp increase of several order of magnitude 
within the first few MeV above threshold, and 
this reflects the GT$^{+}$ distributions. For electron energy $E_e \geq 10$ MeV 
the calculated cross sections display a more gradual increase.  A very similar 
energy dependence of the cross sections is calculated for the neighboring 
even-even parent nuclei $^{48}$Ti and  $^{50}$Cr, in Figs.~\ref{eccs_Ti48}
and ~\ref{eccs_Cr50}, respectively. At low energies all Skyrme HF+RPA 
cross sections are below the values based on SMMC calculations. This 
is especially pronounced in $^{56}$Fe$(e^-,\nu_e)^{56}$Mn, and much 
less in $^{50}$Cr$(e^-,\nu_e)^{50}$V. Cross sections calculated at 
very low electron energies will be very sensitive to the discrete level 
structure of the Gamow-Teller transitions. These cross sections, however, 
are several orders of magnitude smaller than those for $E_e \geq 10$ MeV and, 
when folded with the electron flux to calculate capture rates, 
the differences between values predicted by various models in the low-energy 
interval will not have a pronounced effect on the electron capture rates \cite{Dean.98}. 
Note, however, that in general this will strongly depend on the matter density and 
temperature of the environment. 
More important could be the differences at higher electron energies $E_e > 10$ MeV, 
for which the Skyrme HF+RPA model systematically predicts cross sections above the 
values based on the SMMC. The reason for this systematic effect is most probably 
that SMMC calculations are carried out only in the $0\hbar\omega$ $fp$-shell space. 
Note also that the spread in the calculated cross sections at low energies is greatly 
reduced for higher incident energies and, above $E_e \approx 15$ MeV, all 
Skyrme effective interactions used in the present HF+RPA calculation effectively predict 
a universal behavior of the total electron capture cross sections.


\begin{figure}
\centerline{
\includegraphics[scale=0.6,angle=0]{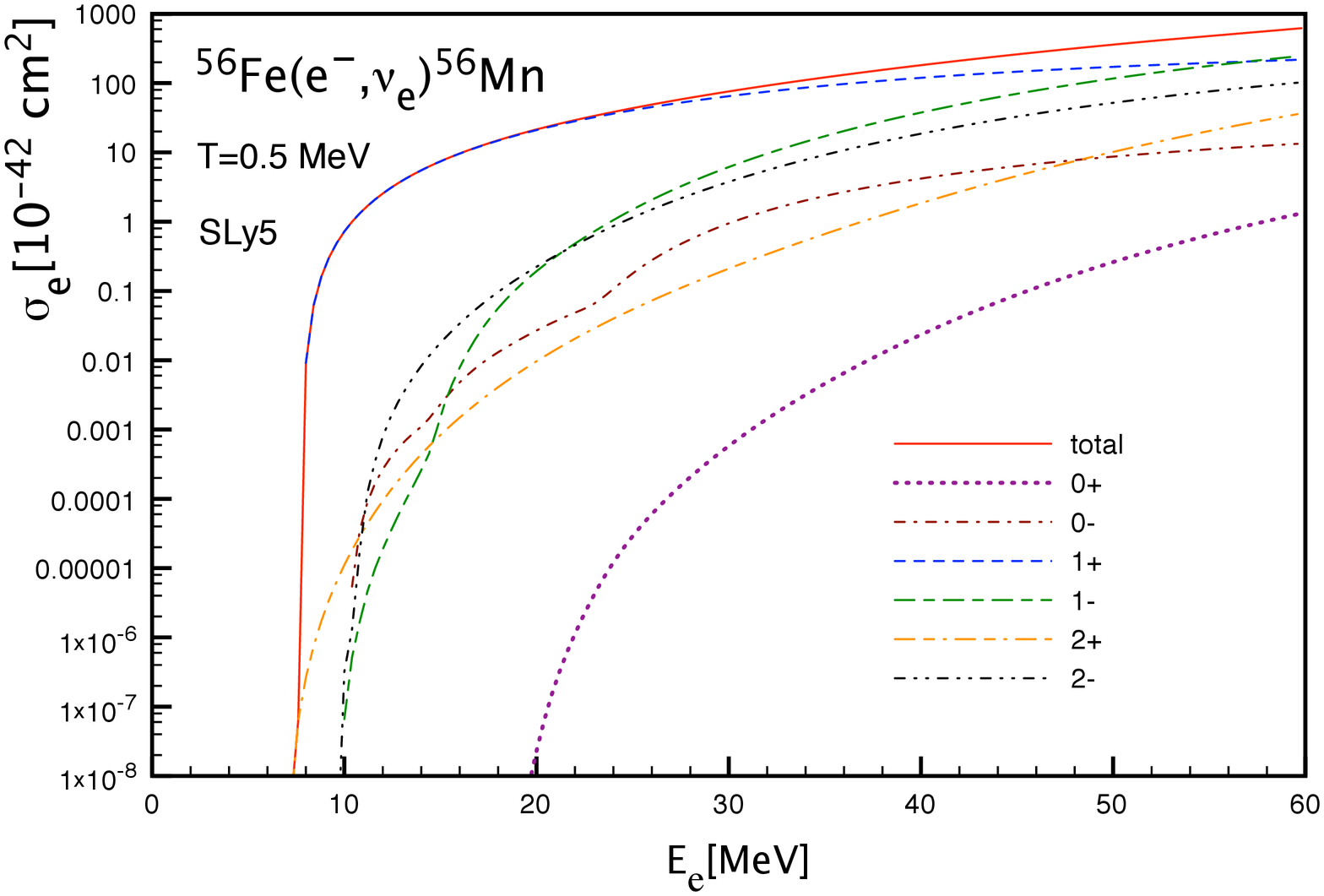}
}
\caption{(Color online) Electron capture cross section for the $^{56}$Fe
target nucleus at $T=0.5$ MeV, calculated 
with the FTHF+FTRPA using the SLy5 Skyrme effective interaction. 
In addition to the total cross section which includes multipole 
transitions $J^{\pi} = 0^{\pm}$, $1^{\pm}$, and $2^{\pm}$, 
contributions from the individual channels are shown in the plot 
as functions of the incident electron energy $E_e$.}
\label{eccs_mult}
\end{figure}

In general, for heavier nuclei and higher electron incident energies, 
not only the $1^+$, but other multipole transitions will also contribute 
to the total cross section for electron capture (cf. the sums in Eq.~(\ref{ec_rate})).
In calculations for the iron group nuclei based on the shell-model, the cross sections 
were calculated only from the $0\hbar\omega$ Gamow-Teller 
strength distributions \cite{Dean.98,LM.00}. In Fig.~\ref{eccs_mult} we plot the 
electron capture cross section for the $^{56}$Fe
target nucleus at T=0.5 MeV, calculated 
with the FTHF+FTRPA using the SLy5 Skyrme effective interaction. 
In addition to the total cross section which includes multipole 
transitions $J^{\pi} = 0^{\pm}$, $1^{\pm}$, and $2^{\pm}$, 
contributions from the individual channels are shown in the plot, 
as functions of the incident electron energy $E_e$. In this case all
the way up to $E_e \approx 30$ MeV the total cross section is completely 
dominated by the $1^+$ channel, with contributions from other 
channels being orders of magnitude smaller. Only at 
very high electron energies contributions from other multipole 
transitions become sizeable. 

\begin{figure}
\centerline{
\includegraphics[scale=0.6,angle=0]{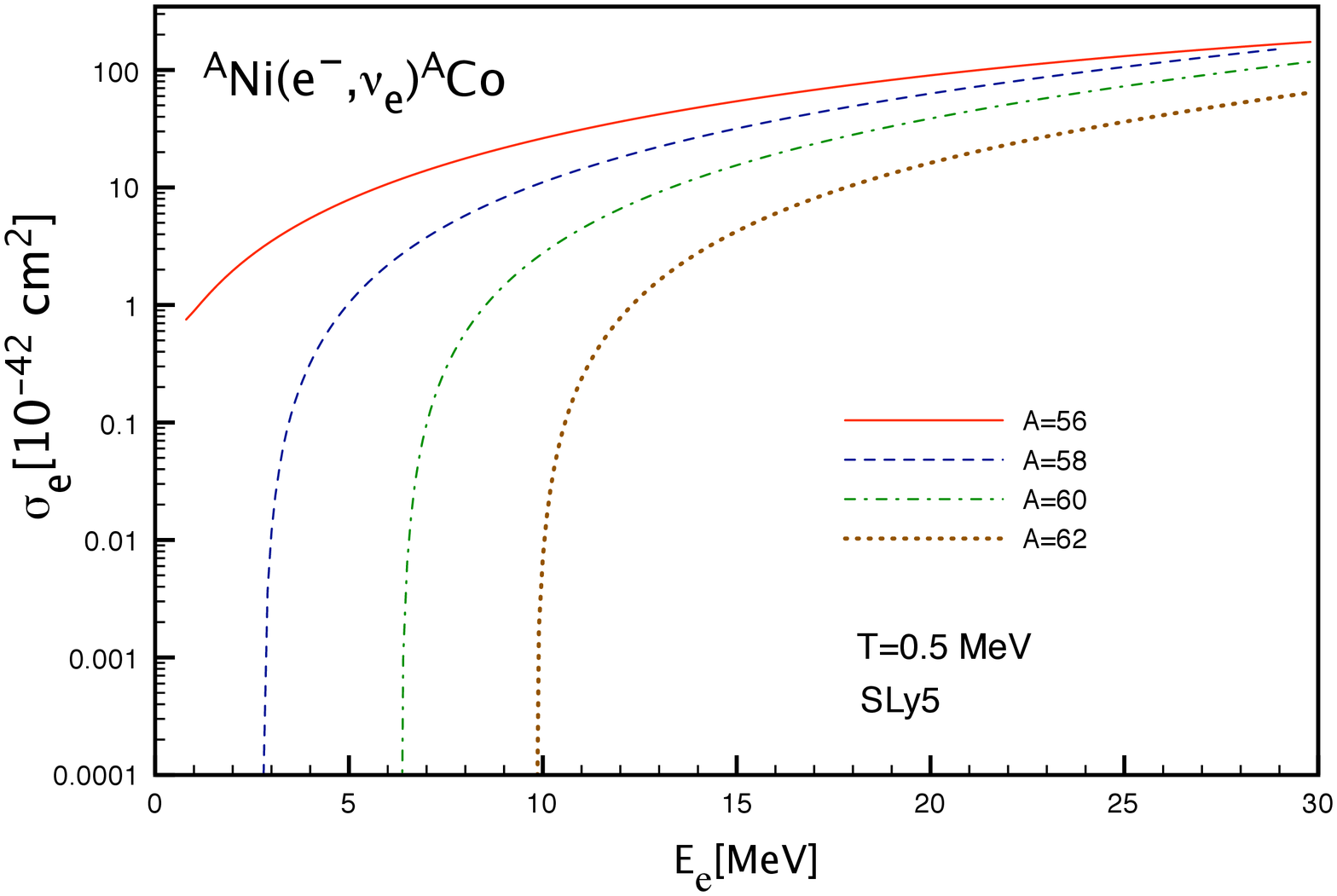}
}
\caption{(Color online) Electron capture cross sections for the even-even Ni 
target nuclei (A=56-62) at $T=0.5$ MeV, calculated 
with the FTHF+FTRPA using the SLy5 Skyrme effective interaction.}
\label{eccs_Ni}
\end{figure}

Finally, in Fig.~\ref{eccs_Ni} we illustrate the isotopic dependence of electron 
capture cross sections. Cross sections for even-even Ni target nuclei, i.e. 
for the reactions $^{A}$Ni$(e^-,\nu_e)^{A}$Co, ($A=56-62$) at T=0.5 MeV, 
are calculated in the FTHF+FTRPA with the Skyrme  SLy5 interaction. At any given 
incident electron energy the calculated cross sections decrease systematically 
along the Ni isotope chain, because more and more neutron orbitals 
become occupied, and therefore not accessible for electron capture reactions. 
Since with the increase of the
number of neutrons the Q value for electron capture increases, more energetic 
electrons are required for the capture reaction on neutron rich isotopes and, 
at any given electron energy, the total  
cross sections are smaller.  We note that the predicted isotopic dependence
of electron capture cross sections in Ni nuclei is in qualitative agreement with 
the results of the SMMC-based study of Ref.~\cite{Dean.98}.


\section{\label{SecGe} Stellar electron capture on neutron-rich Ge isotopes}

At higher densities and temperatures $T \approx 1$ MeV during the collapse phase, 
electrons are also captured on heavier and more neutron-rich nuclei with protons
in the $pf$-shell ($Z < 40$) and neutrons $N \geq 40$. In a 
naive independent particle picture the Gamow-Teller transitions which, 
as it was shown in Sec.~\ref{SecFe}, dominate electron capture in the $pf$-shell, 
are forbidden for nuclei with $Z < 40$ and $N \geq 40$. However, as it has 
been demonstrated in several studies, GT transitions in these nuclei are unblocked 
by finite temperature excitations. A very detailed study based on the random 
phase approximation \cite{CW.83}, has shown that electron capture on
nuclei with protons in the $pf$-shell and $N>40$ can compete with
capture on free protons if forbidden transitions are taken into account,
in addition to allowed ones.  At high temperatures $T \sim 1.5$ MeV 
Gamow-Teller transitions are thermally unblocked  as a result of the 
excitation of neutrons from the $pf$-shell into the $g_{9/2}$ orbital. In 
Ref.~\cite{LKD.01} a hybrid model has been introduced to calculate 
electron capture rates on neutron-rich nuclei in this mass region. 
In the hybrid model the temperature and configuration-mixing effects are 
taken into account with the Shell Model Monte Carlo (SMMC) method,
and are described by partial occupation numbers for the various 
single-particle orbits. Using mean-field wave function with 
finite temperature occupation numbers determined from SMMC,
the electron capture cross sections are calculated with
an RPA approach. Both allowed GT and forbidden transitions 
are included in the calculation. The SMMC/RPA 
hybrid approach was applied to the even germanium isotopes $^{68-76}$Ge 
at typical core collapse temperatures $T \sim 0.5-1.3$ MeV, and it was 
demonstrated that configuration mixing is strong enough to unblock the 
Gamow-Teller transitions at all temperatures relevant to core-collapse
supernovae \cite{LKD.01}. The SMMC/RPA model was also used to calculate rates 
for electron capture on nuclei with mass numbers $A=65-112$, at temperature
and densities characteristic for core collapse \cite{Lan.03}. It was shown that 
electron capture on nuclei dominates over capture on free protons, and 
simulations of core collapse have demonstrated that these capture rates 
produce a strong effect on the core collapse trajectory and the properties of 
the core at bounce.

\begin{figure}
\centerline{
\includegraphics[scale=0.6,angle=0]{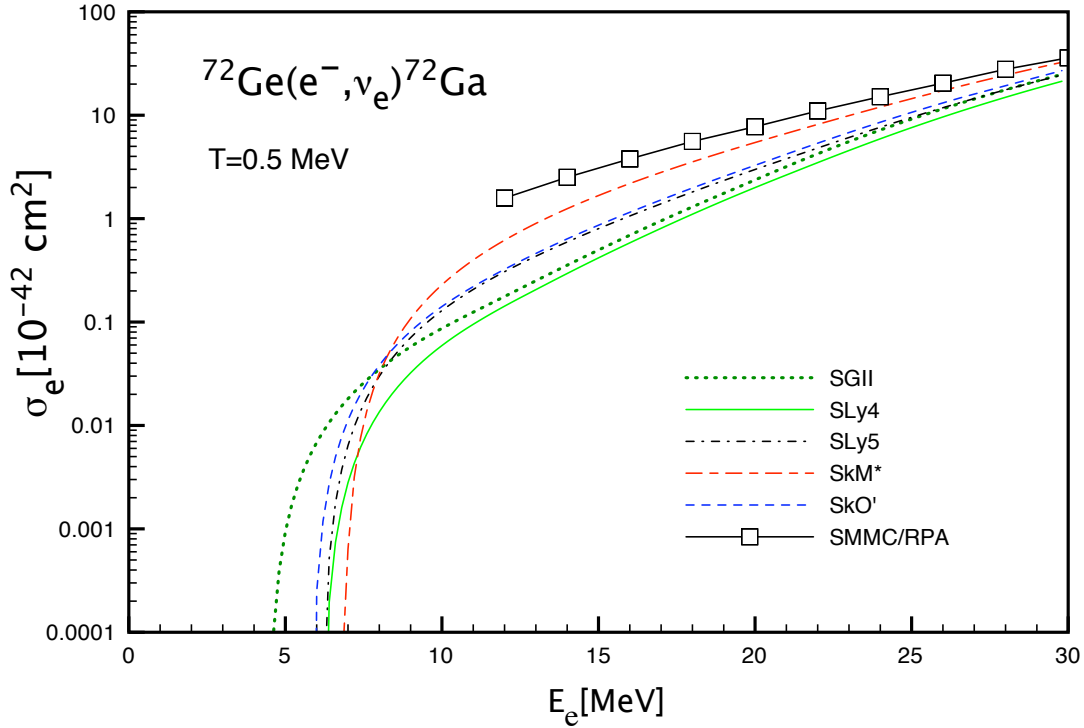}
}
\caption{(Color online) Electron capture cross sections for the reaction
$^{72}$Ge$(e^-,\nu_e)^{72}$Ga at T=0.5 MeV, as functions 
of the incident electron energy $E_e$. The 
Skyrme HF+RPA results are compared with cross 
sections calculated with the hybrid SMMC/RPA model \cite{LKD.01}.}
\label{eccs_Ge72_T0_5}
\end{figure}

In the present analysis we apply the self-consistent finite-temperature Skyrme HF+RPA 
model in the calculation of stellar electron-capture cross sections on neutron-rich 
Ge nuclei, and compare the results with those obtained by Langanke, Kolbe and 
Dean, using the hybrid SMMC/RPA model \cite{LKD.01}.  
In Fig.~\ref{eccs_Ge72_T0_5} we plot the various Skyrme HF+RPA  electron 
capture cross sections for the reaction $^{72}$Ge$(e^-,\nu_e)^{72}$Ga at T=0.5 MeV, 
as functions of the incident electron energy $E_e$, in comparison with the 
SMMC/RPA results. As in the calculation for the iron group nuclei in Sec.~\ref{SecFe}, 
the set of Skyrme effective interactions includes: SGII \cite{gia81}, SkM* \cite{bar82}, 
SLy4\cite{Cha.97}, SLy5\cite{Cha.98},  and SkO' \cite{Rei.99}. A given interaction 
is consistently used both in the finite temperature SHF equations that determine 
the single-nucleon basis, and in the matrix equations of the 
finite temperature RPA. The SMMC calculation of 
Ref.~\cite{LKD.01} included the complete $(pfg_{9/2})$ shell-model space 
and used a pairing+quadrupole residual
interaction with parameters adjusted for this mass region. The single-particle
energies were adopted from the KB3 interaction, but the
$f_{5/2}$ orbital was artificially reduced  by 1 MeV to simulate the effects of the
$\sigma \tau$ component that is missing in the residual interaction.
An energy splitting of 3 MeV between the $g_{9/2}$ and the
$f_{5/2}$ orbitals was assumed. For the RPA calculation based on SMMC, 
the single-particle energies 
were taken from a Woods-Saxon parameterization, and the residual interaction is
a finite-range G-matrix derived from the Bonn nucleon-nucleon potential \cite{Kol.99}.
In the calculation of total cross sections, 
both models include the multipole transitions: $J^{\pi}=0^{\pm},1^{\pm}$, and $2^{\pm}$.  
In the SMMC/RPA calculation the GT strength is quenched by multiplying the GT 
transition matrix elements by the constant factor 0.7. In the present analysis the 
standard reduction of the axial vector coupling constant is employed $g_A = 1.262$ 
$\to$  $g_A = 1.0$, which corresponds to the quenching factor 0.8.
\begin{figure}
\centerline{
\includegraphics[scale=0.6,angle=0]{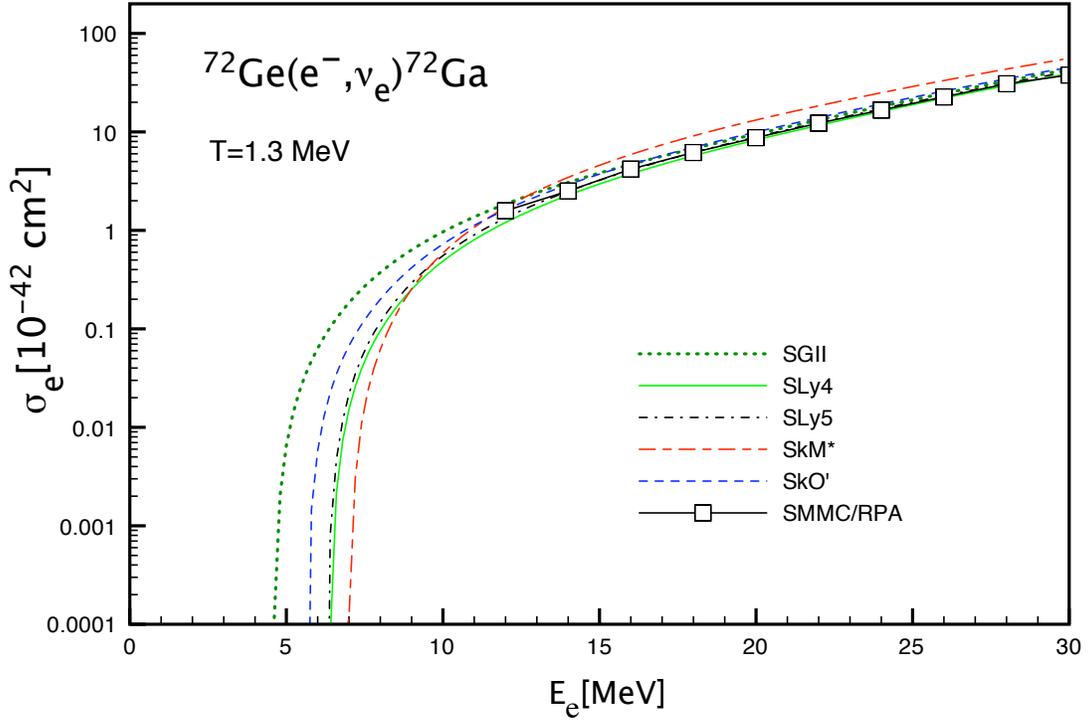}
}
\caption{(Color online) Same as Fig.~\ref{eccs_Ge72_T0_5}, but for 
the temperature T = 1.3 MeV.}
\label{eccs_Ge72_T1_3}
\end{figure}

Similar to the case of iron group nuclei, the cross sections calculated
with different Skyrme parameterizations display a spread of values of 
less than an order of magnitude at lower electron energies.  At higher 
incident energies the differences between values calculated with 
different effective interactions are much smaller. In general the 
Skyrme FTHF+FTRPA results are in good agreement with the cross 
sections calculated in the hybrid SMMC/RPA model, especially at 
higher electron energies $E_e > 20$ MeV. At lower energies the 
SMMC/RPA cross sections are considerably above the results obtained 
in the present calculation. The Skyrme FTHF+FTRPA calculations have also been
carried out at higher temperature: $T=1.3$ MeV (Fig.~\ref{eccs_Ge72_T1_3}). 
It is interesting to note that in this particular case the FTHF+FTRPA results are
indeed very close to those obtained with the SMMC/RPA model.
Note that in Ref.~\cite{LKD.01} values of the calculated cross sections are 
only reported for incident electron energies $E_e > 12$ MeV.
\begin{figure}
\centerline{
\includegraphics[scale=0.6,angle=0]{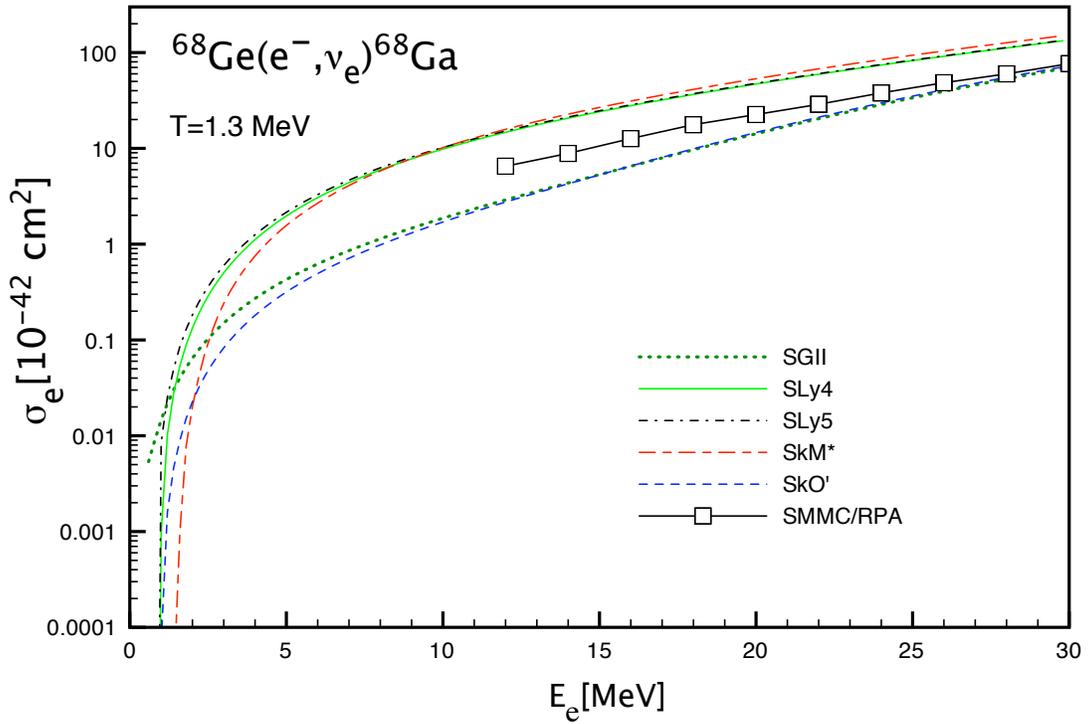}
}
\caption{(Color online) Same as Fig.~\ref{eccs_Ge72_T0_5}, but for $^{68}$Ge at T = 1.3 MeV.}
\label{eccs_Ge68_T1_3}
\end{figure}
\begin{figure}
\centerline{
\includegraphics[scale=0.6,angle=0]{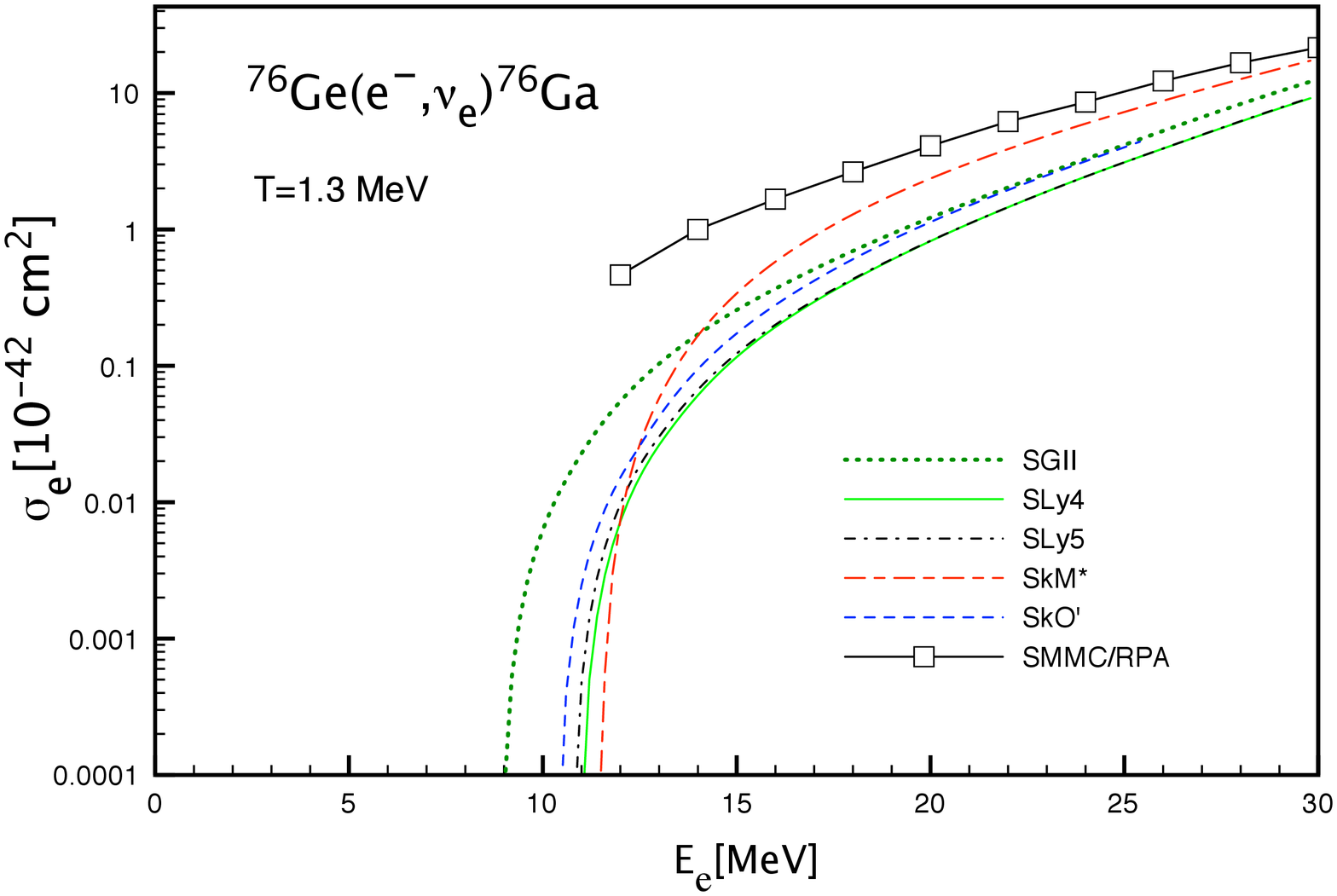}
}
\caption{(Color online) Same as Fig.~\ref{eccs_Ge72_T0_5}, but for $^{76}$Ge at T = 1.3 MeV.}
\label{eccs_Ge76_T1_3}
\end{figure}

In Figs.~\ref{eccs_Ge68_T1_3} and \ref{eccs_Ge76_T1_3} the electron
capture cross sections are shown for the $^{68}$Ge and $^{76}$Ge target nuclei at
$T=1.3$ MeV, respectively.  The Skyrme FTHF+FTRPA results are in qualitative 
agreement with the values calculated in the SMMC/RPA model. For $^{68}$Ge, 
in particular, the Skyrme interactions divide into two branches: SLy4, SLy5 and 
SkM* predict cross sections that are systematically larger than those obtained  
from the SMMC/RPA model, whereas cross sections calculated with SGII and SkO' 
are below the SMMC/RPA results for electron energies $E_e < 30$ MeV. 
The Skyrme FTHF+FTRPA cross sections are systematically smaller than the 
values predicted by the hybrid SMMC/RPA model for the target nucleus $^{76}$Ge. 
The isotopic dependence of the electron capture cross sections, illustrated in 
Fig.~\ref{eccs_Ni} for the even-even Ni nuclei, is also observed in 
Figs.~\ref{eccs_Ge72_T1_3} - \ref{eccs_Ge76_T1_3} for the Ge isotopes. 
With the increase of the neutron number the threshold for electron capture 
is shifted toward higher electron energies, reflecting the change in the Q-value. 
For a given electron incident energy, the total cross section is reduced with 
the increase of the number of neutrons. 

\begin{figure}
\centerline{
\includegraphics[scale=0.6,angle=0]{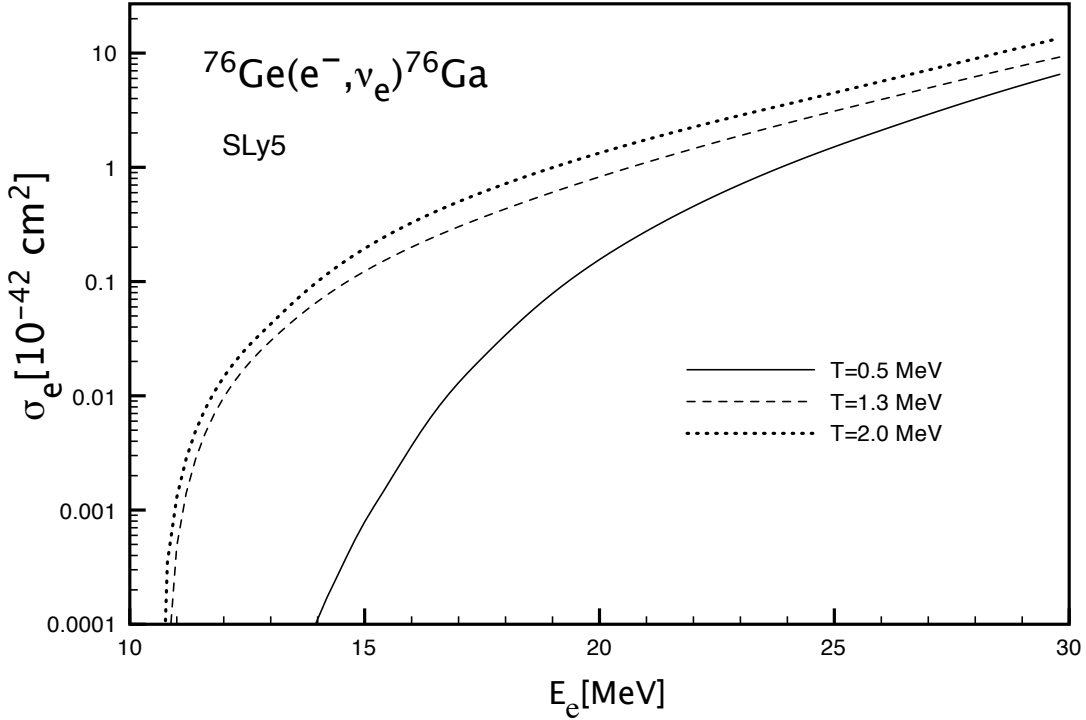}
}
\caption{Electron capture cross sections for the $^{76}$Ge
target nucleus at temperatures $T = 0.5, 1.3$, and 2.0  MeV, calculated 
with the FTHF+FTRPA using the SLy5 Skyrme effective interaction.}
\label{figtemp}
\end{figure}

Finally, in Fig.~\ref{figtemp} we illustrate the temperature dependence of 
electron capture on $^{76}$Ge. The cross sections are calculated with 
the FTHF+FTRPA model at three temperatures: $T = 0.5, 1.3$, and 2.0  MeV, 
using the SLy5 parameterization. The notable increase in the 
calculated cross sections occurs between $T = 0.5$ and 1.3 MeV, and this 
corresponds to a significant thermal unblocking of the neutron 
p$_{3/2}$, f$_{5/2}$, p$_{1/2}$ orbitals (cf. Figs.~\ref{protsub} and \ref{neutsub} 
for the case of $^{74}$Ge). Because these orbitals are already unblocked at 
$T=1.3$ MeV, a further increase in temperature to $T=2.0$ MeV results only in 
a moderate enhancement of the electron-capture cross sections. 

\section{\label{secV} Summary and Conclusions}

Recent advances in modeling nuclear structure phenomena 
have also had a strong impact on astrophysical applications. More and 
more often calculations of stellar nucleosynthesis, nuclear aspects of 
supernova collapse and explosion, and neutrino-induced reactions, 
are based on microscopic global predictions for the nuclear 
ingredients, rather than on simplified semi-empirical approaches. 
In general, the required nuclear input includes properties of 
hundreds of nuclei at and far from the line of $\beta$-stability, including the 
characteristics of strong, electromagnetic and weak interaction 
processes. Many of these nuclei, especially on the neutron-rich 
side, are not accessible in experiments and, therefore, 
nuclear astrophysics calculations crucially depend on accurate 
theoretical predictions for the nuclear masses, bulk properties, 
nuclear excitations, ($n,\gamma$) and ($\gamma,n$) rates, 
$\alpha$- and $\beta$-decay half-lives, fission probabilities, 
electron and neutrino capture rates, etc. 

Improved microscopic stellar weak-interaction rates, evaluated with  
large-scale shell-model diagonalization and/or hybrid RPA models, 
have been employed in recent studies of pre-supernova evolution of massive 
stars, and it has been shown that the resulting changes in the lepton-to-baryon 
ratio and iron core mass lead to significant changes in the hydrodynamics of 
core collapse and the supernova explosion mechanism. These results have 
emphasized the need for accurate microscopic evaluations of nuclear 
weak-interaction rates, at densities and temperatures characteristic for core 
collapse, that can be extended over arbitrary mass regions of the nuclide chart. 
In this work for the first time a self-consistent microscopic framework for 
calculation of weak-interaction rates at finite temperature has been 
introduced, based on Skyrme functionals. 
Single-nucleon levels,  wave functions, and thermal occupation factors for the initial 
nuclear state are determined in the finite-temperature Skyrme Hartree-Fock model, and 
transitions to excited states are computed using the corresponding 
finite-temperature charge-exchange RPA.  Effective interactions are implemented 
self-consistently, i.e.  both the finite-temperature single-nucleon Hartree-Fock equations 
and the matrix equations of RPA are based on the same Skyrme energy density functional. 

The model has been employed in illustrative calculations  for stellar electron 
capture on selected nuclei in the iron group mass region, and for neutron-rich 
isotopes of germanium. Electron-capture cross sections have been calculated as 
functions of the energy of the incident electron, for a representative set of Skyrme 
functionals. By using different Skyrme functionals one is able to estimate the 
range of theoretical uncertainty of the Hartree-Fock plus RPA approach.
For the iron group nuclei, the results have been compared 
with those of Ref.~\cite{Dean.98}, where the SMMC was used to calculate 
Gamow-Teller GT$^{+}$ strength distributions, and electron-capture cross sections 
and rates were computed in the zero-momentum transfer limit. 
At low incident electron energies, at which the cross sections are sensitive to the discrete 
level structure of the Gamow-Teller transitions, all Skyrme HF+RPA 
cross sections are smaller than the values based on SMMC calculations. 
These cross sections, however, are very small and the differences between various models  
will not have a pronounced effect on the calculated electron capture rates. 
More important could be the differences at higher electron energies $E_e > 10$ MeV, 
for which the Skyrme HF+RPA model systematically predicts cross sections larger 
than the values evaluated with the SMMC. It has to be emphasized that the RPA 
approach takes into account large configuration spaces so that for any multipole 
operator the whole sum rule is exhausted, whereas generally this is not the case 
in shell-model calculations. 

For electron capture on neutron-rich Ge nuclei, the finite-temperature 
Skyrme Hartree-Fock plus RPA cross sections have been analyzed in comparison 
with results obtained using the hybrid SMMC/RPA model \cite{LKD.01}, in which 
the nucleus is described as a Slater determinant with thermal occupation numbers 
determined with the SMMC, and capture rates are computed using a charge-exchange 
RPA built on top of the temperature-dependent Slater determinant. In general, a very 
good agreement has been found between the Skyrme FTHF+FTRPA results and the 
cross sections calculated in the hybrid SMMC/RPA model, especially at 
higher electron energies $E_e > 20$ MeV, and higher temperatures $T > 1$ MeV. 
In all cases the two models predict a similar dependence of the cross sections 
on electron energy  in the interval  $12 \leq E_e \leq 30$ MeV. There are, however, 
differences in the absolute values, especially at relatively low temperature ($T=0.5$ MeV), 
electron energies $E_e < 20$ MeV, and for heavier isotopes, e.g. $^{76}$Ge. 

The results of the present study show that the finite-temperature Skyrme 
Hartree-Fock plus charge-exchange RPA framework provides a valuable 
universal tool for the evaluation of stellar weak-interaction rates. Based on a universal 
Skyrme energy density functional, in the sense that the same functional is used for 
all nuclei, this framework can be employed in studies of weak-interaction processes 
in different mass regions. At relevant incident electron energies 
the absolute spread in the electron-capture cross sections, computed with a variety 
of Skyrme functionals, is less than an order of magnitude. The next step will be a 
more extensive calculation and tabulation of electron capture rates for nuclei in 
the mass range $A \approx 60-80$,  which can be compared with modern 
semi-empirical  estimates of weak-interaction rates for intermediate mass nuclei, 
calculated using available experimental information and simple estimates for 
the strength distributions and transition matrix elements \cite{PF.03}. 
For open-shell nuclei at very low temperatures, and especially for calculation 
of $\beta$-decay rates, the framework could be extended to include pairing 
correlations. In the zero-temperature limit, isobaric analog states and GT 
resonances in open-shell nuclei have recently been studied with the newly developed  
self-consistent quasiparticle charge-exchange RPA based on
Skyrme functionals~\cite{Fra.05,Fra.07}. Another interesting 
extension of the model would be the use of a different class of nuclear energy 
density functionals, for instance relativistic EDFs, in which case excited states 
could be calculated using the charge-exchange relativistic QRPA \cite{Paar2004}.

\bigskip\bigskip
This work was supported in part by the Unity through 
Knowledge Fund (UKF Grant No. 17/08), the MZOS - project 1191005-1010, 
and by NExEN ANR-07-BLAN-0256-01.

%

\newpage

%

\end{document}